\documentclass[nofootinbib]{revtex4}
\usepackage[english]{babel}
\usepackage{array,booktabs}   
\usepackage{array} 
\usepackage{amsmath} 
\usepackage{relsize}
\usepackage{wrapfig}
\usepackage{calc}
\usepackage{pdflscape}
\usepackage{color}
\usepackage{float}
\usepackage[font=small,labelfont=bf]{caption}
\usepackage{graphicx}
\usepackage{amsmath}
\usepackage[nodisplayskipstretch]{setspace}

\usepackage{setspace}
\usepackage{tabularx}

\usepackage{float}
\usepackage{color}
\usepackage{amsmath}
\usepackage{float}
\usepackage{calc}
\usepackage{pdflscape}
\usepackage{color}
\usepackage{float}
\usepackage{graphics}
\usepackage{epsfig}
\usepackage{epstopdf}
\usepackage{amssymb}
\usepackage[font=small,labelfont=bf]{caption}
\usepackage{graphicx}
\usepackage{epstopdf}
\usepackage{appendix}
\usepackage{romannum}
\usepackage{soul}
\usepackage{color}

\definecolor{blizzardblue}{rgb}{0.67, 0.9, 0.93}
\definecolor{bubblegum}{rgb}{0.99, 0.76, 0.8}
\usepackage[urlcolor=blizzardblue]{hyperref}
\hypersetup{
    colorlinks=false,
    linkcolor=blue,
    filecolor=magenta,      
    urlcolor=bubblegum,
}
\begin{document} 

\title{Impact of Cross-Sectional Uncertainties on DUNE Sensitivity due to Nuclear Effects.}
\author{ Srishti Nagu$^{1}$ \footnote{E-mail: srishtinagu19@gmail.com},Jaydip Singh$^{1}$ \footnote{E-mail: jaydip.singh@gmail.com}, Jyotsna Singh$^{1}$ \footnote{E-mail: singh.jyotsnalu@gmail.com}, 
R.B. Singh$^{1}$ \footnote{Email: rajendrasinghrb@gmail.com}} 

\affiliation{Department Of Physics, University of Lucknow, Lucknow, India.$^{1}$}

\begin{abstract}
 In neutrino oscillation experiments precise measurement of neutrino oscillation parameters is of prime importance as well as a challenge.  To improve the statistics, presently running 
 and proposed experiments are using heavy nuclear targets. These targets introduce nuclear effects and the quantification of these effects on neutrino oscillation 
 parameters will be decisive in the prediction of neutrino oscillation physics. Limited understanding of neutrino nucleus interactions and inaccurate reconstruction of neutrino energy 
 causes uncertainty in the cross section. The error in the determination of cross section which contributes to systematic error introduces error in the neutrino mixing parameters that are 
 determined by these experiments. In this work we focus on the variation in the predictions of DUNE potential, arising due to systematic uncertainties, using two different event generators-GENIE and GiBUU.
 These generators have different and independent cross-section models. To check the DUNE potential with the two generators we have checked the sensitivity studies of DUNE for CP violation,
 mass hierarchy and octant degeneracy.

\end{abstract}

\maketitle

\section {Introduction}	
The aim of studying the properties and interactions of the most elusive particles neutrinos is to infer their true nature and explore physics beyond the standard model. The past decades have witnessed 
remarkable discoveries in the field of neutrino oscillation physics with the help of phenomenal experiments, substantiating the existence of neutrino oscillations. Neutrino oscillation implies the
change of neutrino flavor as they travel i.e., a neutrino which is generated with a certain flavor after traveling a certain distance might end up having a different flavor. Neutrino oscillation
parameters that govern neutrino oscillation physics are mixing angles $\theta_{ij}$ where $j>i$=1,2,3 ($\theta_{12}, \theta_{13}, \theta_{23}$), dirac phase $\delta_{CP}$ and the magnitude of mass
squared differences, $\Delta m_{21}^2$ known as solar mass splitting and $\Delta m_{31}^2$ known as atmospheric mass splitting. Much progress on the precise determination of neutrino oscillation 
parameters has been made by achieving nearly precise determination of the mixing angles $\theta_{12}$, $\theta_{23}$ and non-zero value of $\theta_{13}$ \cite{theta13_1,theta13_2,theta13_3} and mass
squared differences $\Delta m_{21}^2$, $|\Delta m_{31}^2|$. The remaining unknown parameters on the canvas of neutrino oscillation physics are- (\romannum1)the sign of $\Delta m_{31}^2$ or the neutrino 
mass ordering. There are two possibilities of arrangement, for the neutrino mass eigenstates $m_{i}$(i=1,2,3). One is normal mass ordering or the normal mass hierarchy(NH) where the neutrino mass 
order is- $m_{1}\ll m_{2} \ll m_{3}$ and the other is inverted mass ordering or the inverted mass hierarchy(IH) where the neutrino mass order is- $m_{2}\approx m_{1} \gg m_{3}$  (\romannum2) determination
of the octant of $\theta_{23}$, whether the value of $\theta_{23}$ lies in the lower octant(LO) $0< \theta_{23} < \pi/4$ or higher octant(HO) $\pi/4 < \theta_{23} < \pi/2$. This uncertainty in the octant
of $\theta_{23}$ is known as the octant degeneracy problem which describes the incapability of an experiment to distinguish between the values $\theta_{23}$ and $(\pi/2-\theta_{23})$ (\romannum3) determination
of the value of dirac phase $\delta_{CP}$ which can lie in the range $-\pi < \delta_{CP} < \pi$. As we know if the value of this parameter differs from 0 or $\pi$, it would indicate CP violation in the
leptonic sector. This discovery can shed light on the origin of leptogenesis \cite{leptog}  and can be a tool to answer some of the intriguing questions like baryon asymmetry of the universe \cite{baryog}. 
Precise CP phase value is also required for the exact absolute neutrino mass measurement in double beta decay experiments and also for explaining the sterile neutrinos phenomenon \cite{jd}. 
The global analysis \cite{0c} shows two sets of best fit values of neutrino oscillation parameters in 1$\sigma$ and 3$\sigma$ ranges that correspond to the analysis done, with and without Super-Kamiokande 
atmospheric neutrino data.

The pre-requisite for precise knowledge of neutrino oscillation physics depends on many factors amongst which precise reconstruction of neutrino energy is of extreme importance. As we know that the neutrino
oscillation probability itself depends on the energy of the neutrinos, any incorrect measurement of neutrino energy will be propagated to the measurements of neutrino oscillation parameters since it causes 
uncertainties in the cross section measurement and event identification. 
Many important long baseline neutrino oscillation experiments use accelerator generated neutrino beams. These neutrino beams are not monoenergetic, thus for the reconstruction of neutrino energy complete
information of final state particles is required. The energy reconstruction of neutrinos from final state particles need careful examination since the identification of final state particles in presence of
nuclear effects is a challenging task because the particles produced at the initial neutrino-nucleon interaction vertex and the particles captured by the detector can be different or not identical.
Presently running and proposed future experiments use heavy nuclear targets in order to collect large event statistics as a \textit{a priori} requirement of neutrino oscillation experiments but at the same
time, the use of heavy nuclear targets gives a boost to nuclear effects. The high event statistics obtained, minimize the statistical error and shifts the attention to explore a handle to control systematic errors. 
The uncertainties in the determination of neutrino-nucleus cross-sections arising due to the presence of nuclear effects are one of the most important sources of systematic errors. It is important to investigate
the precise neutrino-nucleon interaction cross-sections in an attempt to reduce systematic errors. The studies of interrelation between uncertainties in neutrino-nucleon 
cross sections and its impact on the determination of neutrino oscillation parameters have been explored previously in many research works \cite{crossoss1,crossoss2,crossoss3,crossoss4}. 
To control the systematic errors, the current knowledge of nuclear effects is still insufficient, as stated in \cite{nuceffect1,nuceffect2,nuceffect3,nuceffect4,snaaz,moselcross}.

Here, in an attempt to capture nuclear effects, we have selected two different simulation tools, GENIE \cite{4} and GiBUU \cite{5}. Both the neutrino event generators incorporate nuclear effects in their simulation
codes but differ in the selection of nuclear models and computation of various neutrino-nucleus interaction processes. The nucleus is a collection of nucleons and the study of the effect of all the nucleons in neutrino-
nucleus interactions is not trivial. Different neutrino event generators which include nuclear effects in their analysis program use different approximations to define nuclear effects. 
Since the result of an experiment must be model independent, this motivated us to perform our analysis.

The future Long-Baseline Deep Underground Neutrino Experiment(DUNE) \cite{ref0,ref1,ref4,ref2,ref3}, as one of the third generation neutrino experiment is in the process of being set up at United States and
aims to explore the key problems in neutrino physics i.e. the determination of neutrino mass hierarchy, octant degeneracy, CP violation and certainly new physics. Pinning down the systematic uncertainties 
in the proposed experiment will help us to achieve these goals up to the desired sensitivity. An intense megawatt scale muon neutrino beam produced at Fermilab will aim at two detectors: 
the near detector, 575 meters from the target (fine-grained magnetic spectrometer) and the far detector which will be a suite of four independent 10 kt Liquid Argon Time Projection Chambers(LArTPCs), 
situated 1300 km from the target in the Sanford Underground Research Facility(SURF), South Dakota. The DUNE-LBNF flux spreads in the energy range 0.5 to 10 GeV, having an average energy of 2.5 GeV. 
It is composed of QE(Quasi-Elastic), RES(Resonance), DIS(Deep Inelastic Scattering) and Coherent neutrino-nucleon interaction processes each having a different energy dependent cross-section.

The paper is organized in the following sections: In Section II we describe the neutrino event generators GENIE and GiBUU used in this work including a detailed comparison of the physics incorporated in them.
We outline the simulation and experimental details in Section III, followed by a discussion of the CP sensitivity, mass hierarchy and octant sensitivity results in Section IV. Finally, we present our conclusions
in Section V.

\section{EVENT GENERATORS: GENIE AND GiBUU}
The interaction cross section($\nu$-Ar) used in this work is computed with two neutrino event generators: GENIE(Generates Events for Neutrino Interaction Experiments)2.12.06 \cite{4} and GiBUU(Giessen Boltzmann-Uehling-Uhlenbeck)v-2019 \cite{5}. We have considered the quasi-elastic(QE), resonance (RES) from $\Delta$ resonant decay and contribution from higher resonances, two particle-two hole(2p2h/MEC) and deep inelastic scattering(DIS) interaction processes. The estimated total cross-section from the two generators is further converted into the input format of GLoBES package. Neutrino cross section as a function of neutrino energy is shown in Figure 1. From both the generators, we observe a difference in the value of cross sections for $\nu_{e}$ and $\nu_{\mu}$ both. Above energy 3 GeV, a difference in the cross-sections of $\nu_{e}$ and $\nu_{\mu}$ is observed if they are computed via GiBUU whereas no such difference is observed if they are estimated using GENIE. The difference in the neutrino-nucleon interaction cross-sections introduced by the two generators results in different event rates as a function of reconstructed neutrino energy, as illustrated in Figure 2. The presented event rate is estimated using GLoBES, by considering normal hierarchy as true hierarchy and $\delta_{CP}= -\pi/2$.

\begin{figure}
 \centering\includegraphics[scale=.44]{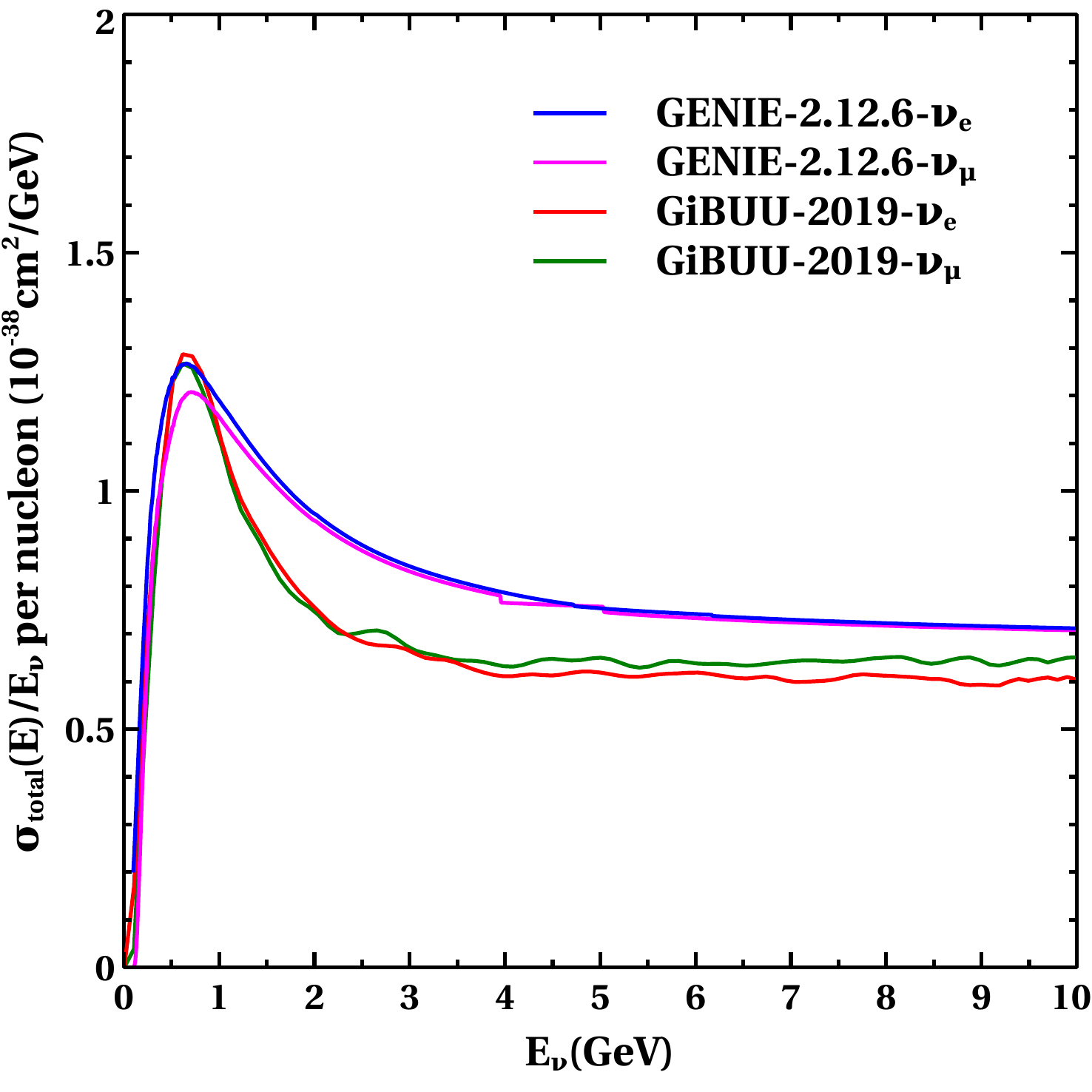}
 \caption{Total neutrino-argon interaction cross-section per nucleon as a function of neutrino energy by GENIE and GiBUU in the energy regime 1-10 GeV, for different charged current processes considered in our work.}
\end{figure}

\begin{figure}
 \centering\includegraphics[scale=.44]{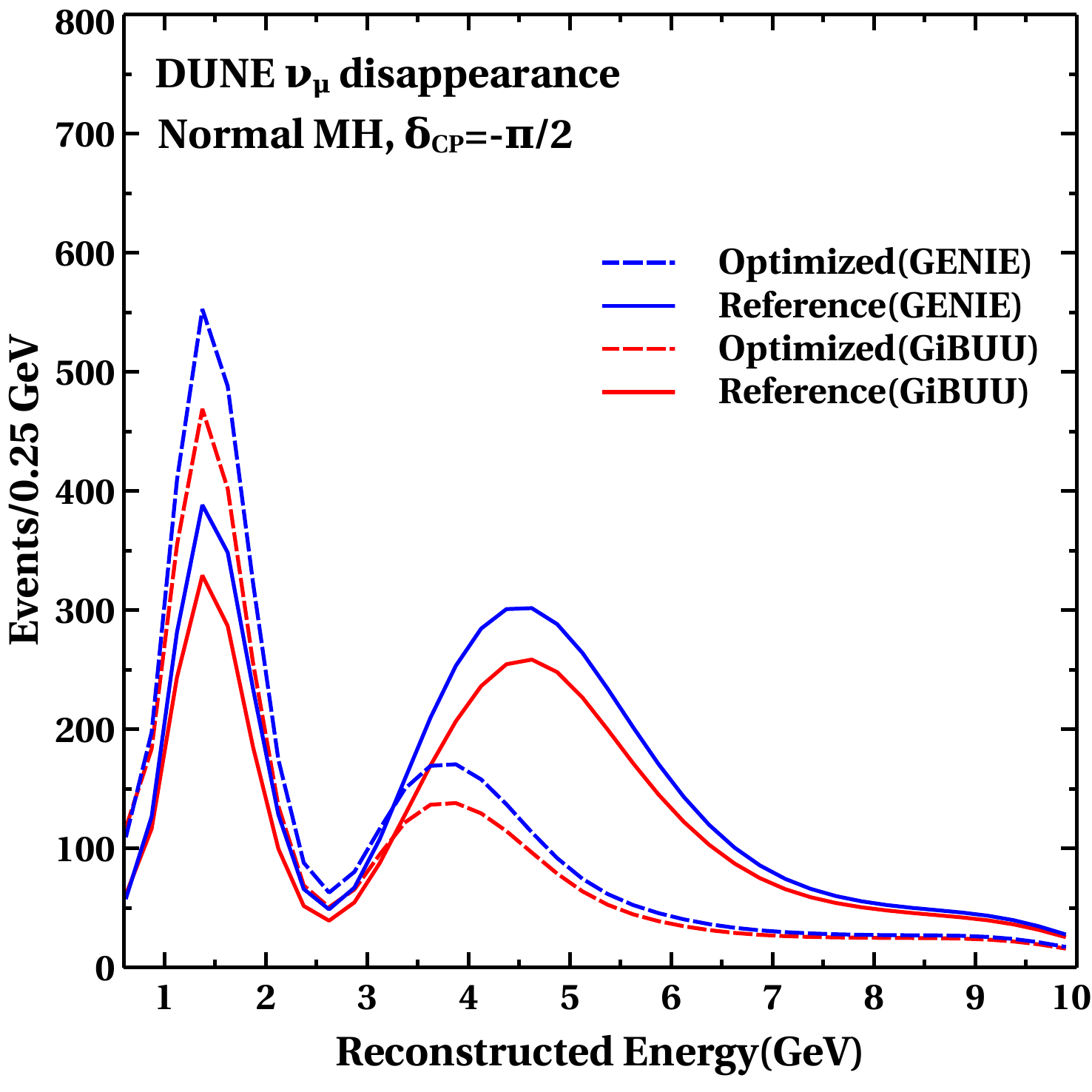}
 \centering\includegraphics[scale=.44]{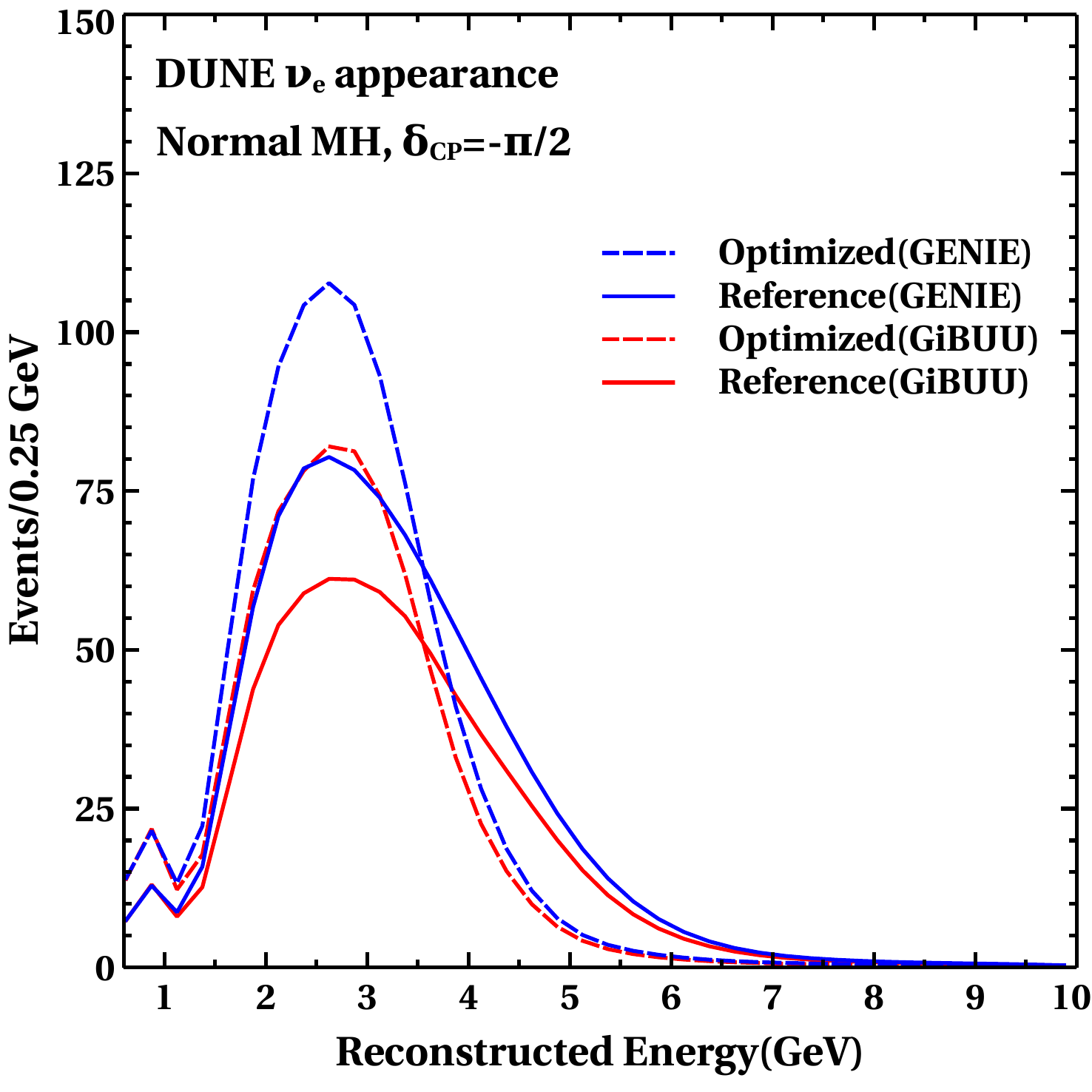}
 \caption{Left panel shows $\nu_{\mu}$ disappearance and right panel shows $\nu_{e}$ appearance event distributions as a function of reconstructed neutrino energy for both reference and optimized beamline designs 
 in the energy regime 1-10 GeV.}
\end{figure}

In this section, we will discuss the qualitative theoretical difference in the nuclear models and the way neutrino-nucleus interaction processes are considered in 
both the generators and some common approach used for the treatment of neutrino-nucleus interaction analysis. The selected event generators differ in the selection of 
nuclear models describing the neutrino-nucleus interactions. GENIE is a ROOT \cite{7a} based neutrino event generator designed using object oriented methodologies and 
developed entirely in C++. It is used by the majority of neutrino baseline experiments running around the world, such as MINERvA \cite{7b}, MINOS \cite{7c}, MicroBooNE \cite{7d}, 
NovA \cite{7e} and T2K \cite{7f}. On the other hand, GiBUU is based on a coupled set of semiclassical kinetic equations, these equations describe the dynamics of a hadronic system
in phase space and time. It is based on FORTRAN routines. GiBUU has been adopted not only by the neutrino community but also by other research communities in nuclear and
particle physics experiments to explain the various interactions viz.($\nu$, e, $\gamma$, A, p, $\pi$)-A, these interaction phenomena has been tuned and explained very well by GiBUU \cite{5,7bb}. 
The RFG model used in GENIE is based on the model suggested by A. Bodek and J.L. Ritche \cite{ABodak}, that includes short-range nucleon-nucleon correlations while in GiBUU, RFG is updated by adding a
density dependent mean-field potential term in which all nucleons are assumed to be bound. When simulating QE neutrino-nucleus interaction process, RFG model \cite{refrfg} is used to describe the nuclear
structure by both the generators. The neutrino-nucleus interaction dynamics is also described by the RFG model \cite{refrfg} under the assumptions of PWIA(Plane Wave Impulse Approximations), further 
details can be found in \cite{refpwia}. Modeling of QE scattering in GENIE is according to Llewellyn Smith model \cite{qemodel}. For details on QE cross section mechanism in GiBUU, one can look 
into \cite{minibooneQE,moselQE}. For nuclear density distribution both the generators use Woods-Saxon parametrization \cite{woodsaxon}. The value of axial mass used by GiBUU is $M_{A}$= 1 GeV/c$^{2}$ 
while GENIE uses a variable value of axial mass between 0.99-1.2 GeV/c$^{2}$. The vector form factors used by GiBUU and GENIE are BBBA07 \cite{refGibuuVct} and BBBA05 \cite{refGenieVct} respectively. 

The average energy of the neutrino beam (LBNF) for the DUNE experiment is $\sim2.5$ GeV and RES is the dominant process at this energy. GiBUU consists of 13 kinds of resonance modes. The MAID 
analysis \cite{refGibuuMAID1,refGibuuMAID2} of the electron scattering data provides vector form factors for each of the 13 resonance modes in GiBUU. On the other hand, GENIE consists of 16 resonance 
modes based on Rein Sehgal model \cite{RShegal}. An additional contribution to the total neutrino cross-section at neutrino energy less than 1 GeV contributes via processes involving two particle
two hole (MEC/2p2h) excitations. This process arises mainly from nucleon-nucleon correlations in the initial state interactions(ISI), neutrino coupling to the 2p2h and FSI. A debate on the importance
of this process can be found in \cite{2p2h}. For simulating DIS processes, GiBUU uses PYTHIA \cite{gibuudis} while GENIE applies the model of Bodek and Yang \cite{bodekyang} along with Aivazis, Olness 
and Tung model \cite{discharm}, particularly for DIS charm production. 

Final state interactions(FSI), significantly modify the event distribution captured by the detectors in the form of changed identities,
topologies and kinematics of the initially produced particles. This necessitates proper modeling of FSI which is incorporated differently in different nuclear models which are in use. 
The treatment of FSI by GENIE and GiBUU is entirely different. GENIE simulates nuclear re-interactions using Intranuke hA and hN as FSI models, details can be found in \cite{dytman1,dytman2}. 
The hA model is data driven while hN incorporates theoretical descriptions. In GiBUU, FSI is modeled by solving the semi-classical BUU equations where various particle species are coupled via 
a mean field potential and collision terms. Further differences in nuclear models, cross section models and FSI models used by the two generators can be found in \cite{ref45}. Recent development 
and its implementation in both the generators i.e. GENIE and GiBUU can be found in \cite{refGenieMEC, refGibuu2p2h}.  
 
\section{SIMULATION AND EXPERIMENTAL DETAILS}
For the simulation of the DUNE experiment, we have considered a far detector with a fiducial volume of 40 kton liquid argon(A=40) placed at a distance of L=1300 km from the wideband 
neutrino beam source with a running time of 3.5 years, each in neutrino and antineutrino mode. The neutrino fluxes used here correspond to the 80 GeV beam configuration \cite{3a}, 
with an assumed beam power of 1.07 MW for two beamline designs (\romannum1) reference design (\romannum2) optimized design. We perform the sensitivity analysis for DUNE with both reference and optimized beams 
to explore the physics potential of DUNE. The main differences between the two beam designs include the geometry of the decay pipe and design of the horn. Further details regarding the potential
beamline designs can be found in \cite{3b} and difference in few parameters are enlisted in Table II.
For performing the sensitivity analysis we have used the GLoBES(General Long Baseline Experiment Simulator) \cite{1,2} package which requires cross-section, neutrino and anti-netrino 
beam fluxes and detector parametrization values as input. The cross-section input format is: $\hat{\sigma}(E) = \sigma(E)/E [10^{-38} cm{^{2}}$/GeV], one can find further details 
in \cite{globesmanual}. The computation of binned event rates is performed by an energy smearing algorithm which we have chosen to be a Gaussian function of energy resolution \cite{globesmanual}. 
The energy resolution for $\nu_{e}$ is 15$\%/\sqrt{E}$(GeV) and for $\nu_{\mu}$ is 20$\%/\sqrt{E}$(GeV) \cite{cadams}. The true values of the oscillation parameters \cite{0c} considered in this
analysis are presented in Table I. The numerical procedure carried out to study the sensitivities is done by calculating $\Delta \chi^{2}$ using the default definition present in GLoBES. 
The oscillation analysis includes both the $\nu_{e}(\overline{\nu_{e}})$ appearance and $\nu_{\mu}(\overline{\nu_{\mu}})$ disappearance channels and the systematics considered in our analysis are
presented in Table III. The relevant background that is considered in this work for the muon disappearance channel is the neutral current interaction.
For the backgound of electron appearance channel we have considered contributions from three different channels i.e. charged current interactions of $\nu_{e} \rightarrow \nu_{e}$, misidentified CC 
$\nu_{\mu} \rightarrow \nu_{\mu}$ and neutral current(NC) backgounds. 
 
\begin{table}[htp]
\caption{Oscillation parameters considered in our work}
\renewcommand\thetable{\Roman{table}}
\centering
\setlength{\tabcolsep}{2pt}
\begin{tabular}{c | c | c }

\hline\hline
  \textbf{Parameter}           &  \textbf{Best Fit Value}      & \textbf{3$\sigma$ Range}\\
\hline
   $\theta_{12}$               &   0.590                  & -                             \\
   $\theta_{13}$               &   0.151                  & -                             \\
   $\theta_{23}$(NH)           &   0.867                  & 0.703 - 0.914   \\
   $\theta_{23}$(IH)           &   0.870                  & 0.710 - 0.917   \\
   $\delta_{CP}$               &   0                           & $-\pi - +\pi$   \\
   $\Delta m^{2}_{21}$         &   7.39e-5 eV$^{2}$     & -                             \\
   $\Delta m^{2}_{31}$(NH)     &  2.525e-3 eV$^{2}$    & +2.427 $\rightarrow$ +2.625    \\
   $\Delta m^{2}_{31}$(IH)     &  -2.512e-3 eV$^{2}$  & −2.611 $\rightarrow$ −2.412    \\
  \hline\hline
\end{tabular}
\end{table}


\begin{table}[htp]
\caption{A comparison of beamline parameters taken from CDR for Reference Design flux and Optimized Design flux are as follows \cite{3b}.
These parameters are used in our analysis.}
\renewcommand\thetable{\Roman{table}}
\centering
\setlength{\tabcolsep}{2pt}
\begin{tabular}{c | c | c }

\hline\hline
\textbf{Parameter}          & \textbf{Reference Beamline Design}  & \textbf{Optimized Beamline Design} \\
\hline  
   Proton Beam Power           & 1.07 MW          & 1.07 MW                \\
   Energy of Proton Beam       & 80 GeV           & 80 GeV                 \\
   Horn Style                  & NuMI-style       & Genetic Optimization   \\
   Horn Current                & 230 kA           & 297 kA                 \\
   Diameter of Decay Pipe      & 4 m              & 4 m                    \\
   Length of Decay Pipe        & 204 m            & 241 m                  \\
\hline\hline
\end{tabular}
\end{table}

\begin{table}[htp]
\caption{Systematic uncertainties for signal and backgound channels used for both the reference and optimized beamline designs.}
\renewcommand\thetable{\Roman{table}}
\centering
\setlength{\tabcolsep}{2pt}
\begin{tabular}{c | c | c }
\hline\hline
   \textbf{Channel}          & \textbf{Signal}  & \textbf{Background} \\
\hline   
   $\nu_{\mu}(\overline{\nu_{\mu}}) \rightarrow \nu_{e}(\overline{\nu_{e}})$      & 5$\%$    & 10$\%$          \\
   $\nu_{\mu}(\overline{\nu_{e}}) \rightarrow \nu_{\mu}(\overline{\nu_{e}})$      & 5$\%$    & 10$\%$          \\
\hline\hline
\end{tabular}
\end{table}

\section{SENSITIVITY STUDIES FOR DUNE}
In this section, we have tried to explore the impact of cross-sectional uncertainties (shortcomings in the theoretical aspect of nuclear physics models as implied in the generators) 
on the three major goals DUNE aims to resolve i.e. (\romannum1)CP phase violation (\romannum2)mass ordering (\romannum3)octant ambiguity. Several attempts have been made to perform the sensitivity analysis for long baseline experiments like DUNE, for example, determination of octant sensitivity and $\delta_{CP}$ sensitivity have been performed in \cite{sens1,sens2,sens3} while a complete sensitivity analysis for other experiments like T2K \cite{7f}, T2HK \cite{t2hk}, NO$\nu$A \cite{7e} along with DUNE have been performed in \cite{sens4,sens5}.

\subsection{CP VIOLATION SENSITIVITY}
In order to observe CP violation, the value of the CP phase must be different from CP conserving values i.e. 0 or $\pm \pi$.
Since we do not know the true value of $\delta_{CP}$, the analysis is performed by scanning all the possible true values of $\delta_{CP}$, over the entire range $-\pi < \delta_{CP} < +\pi$ and
comparing them with the CP conserving values. Our test parameters are $\delta_{CP}, \theta_{23}$ and $|\Delta m^{2}_{31}|$. While performing our analysis we have marginalized over the test parameters in 3$\sigma$
range, as mentioned in Table I. To calculate the CP violation sensitivity we perform calculations as-
\begin{figure}
 \centering\includegraphics[scale=.5]{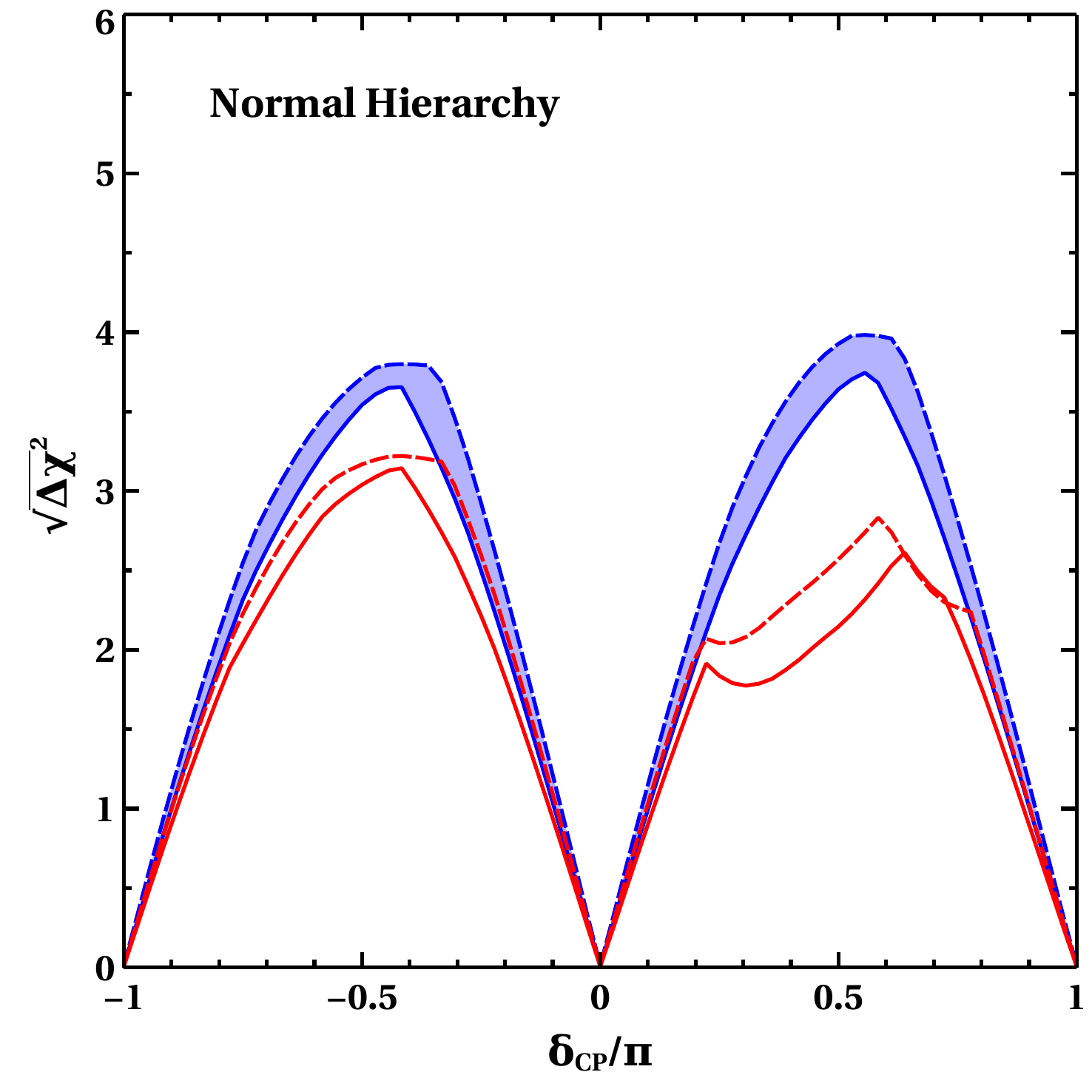}
 \centering\includegraphics[scale=.5]{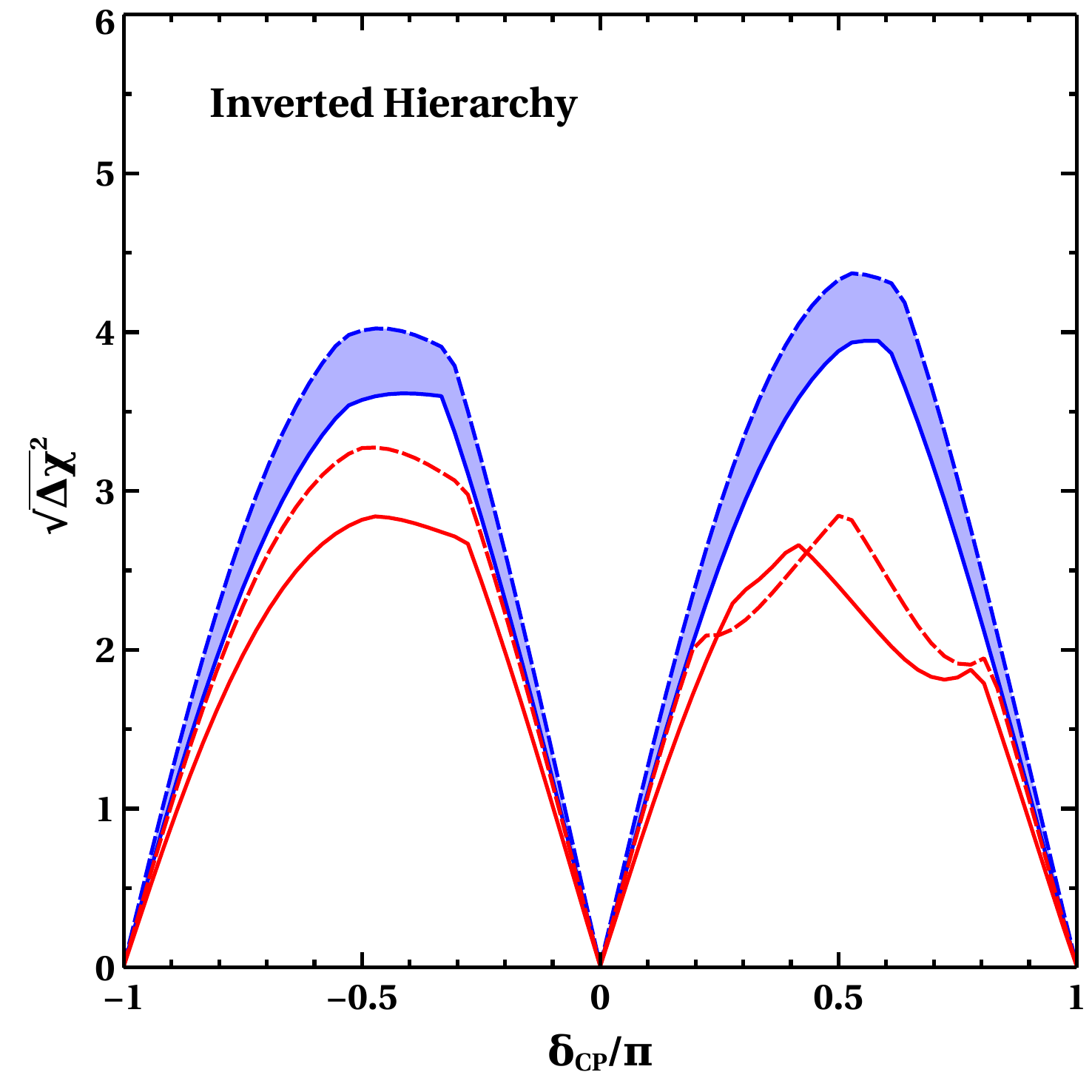}
 \caption{CP sensitivity measurement as a function of true value of $\delta_{CP}$ for NH(left panel) and IH(right panel) by GENIE(blue lines) and GiBUU(red lines). Reference and optimized designs are represented by
 solid and dashed lines respectively.}
\end{figure}

\begin{equation}
 \Delta \chi^{2}_{0} = \chi^{2}(\delta_{CP}=0) - \chi^2_{true}
\end{equation}
\begin{equation}
\Delta \chi^{2}_{\pi} = \chi^{2}(\delta_{CP}=\pi) - \chi^2_{true} 
\end{equation}
\begin{equation}
 \Delta \chi^{2} = min(\Delta \chi^{2}_{0},\Delta \chi^{2}_{\pi})
\end{equation}
Qualitative handle on the measurement of CP violation is obtained by using, $\sigma = \sqrt{\Delta \chi^{2}}$ and is illustrated in Figure 3. Left panel of Figure 3 shows CP sensitivity when normal 
hierarchy is considered as true hierarchy. In this analysis 1$\sigma$ variation is observed in CP sensitivity results at $\delta_{CP} \sim 0.5/\pi$ in the range $0<\delta_{CP}/\pi<1$ 
for the results obtained by GENIE and GiBUU for the DUNE experiment. This gives us a hint that the physics analysis for CP sensitivity from these two generators will have a difference in the results by 1$\sigma$ if mass hierarchy is found to be normal.

This 1$\sigma$ difference in the CP sensitivity is observed for both the reference and optimized beam designs as reflected in Figure 3 left panel.
The right panel of Figure 3 shows CP sensitivity when inverted hierarchy is considered as true hierarchy. The CP sensitivity results with the two different generators show a variation of more than 1$\sigma$ 
in the range $0<\delta_{CP}/\pi<1$ around 0.5/$\pi$ for both reference and optimized beam designs. In the negative half range($-1<\delta_{CP}/\pi<0$) of the CP values, the variation between GENIE and
GiBUU predictions for reference and optimized beam designs is seen to be lesser than 1$\sigma$ for both normal and inverted hierarchy cases.

\subsection{MASS HIERARCHY SENSITIVITY}
Determination of mass hierarchy is one of the most crucial problems in neutrino physics i.e. the quest of finding the true nature of neutrino mass ordering, is normal or inverted.
Mass hierarchy sensitivity is calculated by assuming normal(inverted) hierarchy as true hierarchy and comparing it with inverted(normal) hierarchy by using equation 4(5). 
So we set opposite hierarchies in true and test values. Figure 4 shows the mass hierarchy sensitivity for both the normal hierarchy(left panel) and inverted hierarchy(right panel) cases. 
The $\Delta\chi^{2}$ quantity for mass hierarchy sensitivity, is calculated as below-
\begin{equation}
 \Delta\chi^{2}_{MH} = \chi^{2}_{IH} - \chi^{2}_{NH} 
\end{equation}
\begin{equation}
 \Delta\chi^{2}_{MH} = \chi^{2}_{NH} - \chi^{2}_{IH} 
\end{equation}

\begin{figure}
\centering\includegraphics[scale=.5]{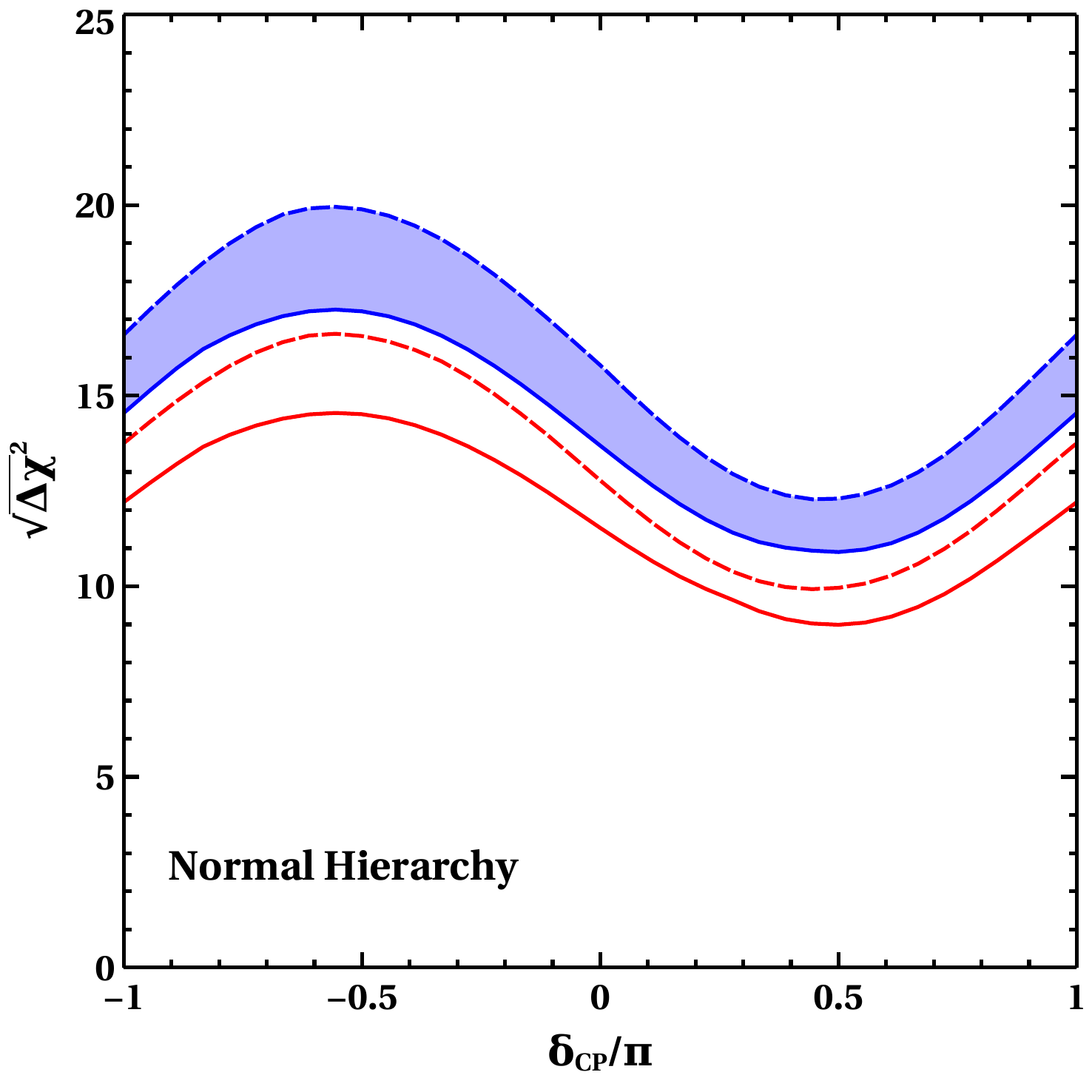}
\centering\includegraphics[scale=.5]{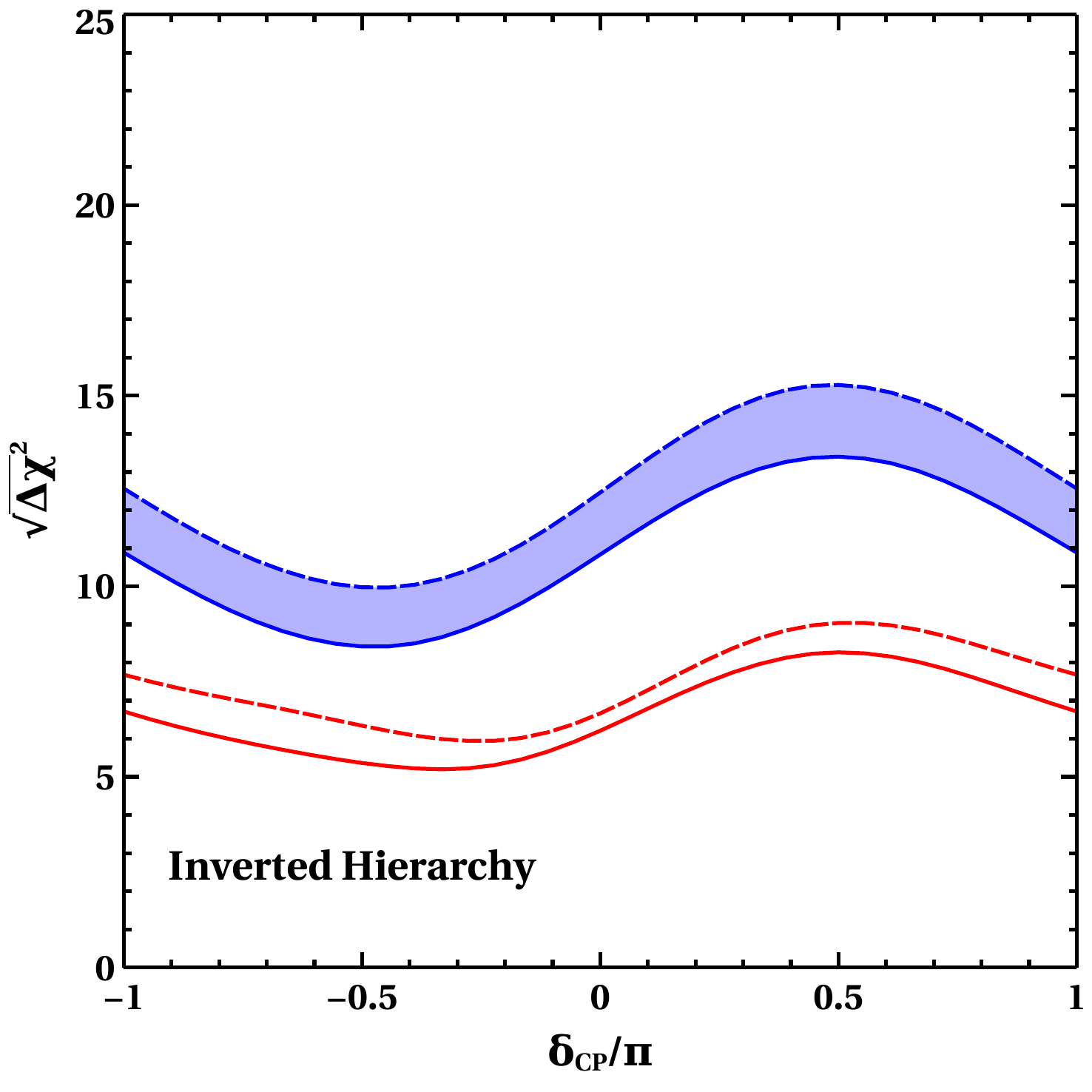}
\caption{Mass hierarchy sensitivity measurement as a function of true value of $\delta_{CP}$ for NH(left panel) and IH(right panel) by GENIE(blue lines) and GiBUU(red lines). Reference and optimized designs 
are represented by solid and dashed lines respectively.}
\end{figure}

For the true normal hierarchy case presented in the left panel of Figure 4, we observe a variation of more than 2$\sigma$ for reference beam and more than 3$\sigma$ for optimized beam predictions made by
GENIE and GiBUU in the range, $-1<\delta_{CP}/\pi<0$ at $\sim -0.5/\pi$ while in the range of positive $\delta_{CP}$ values i.e. $0<\delta_{CP}/\pi<1$ we observe a variation of around 1$\sigma$ at $\sim0.5/\pi$ between the GENIE and GiBUU predictions, for both the beams. This analysis illustrates that there will be a difference in the sensitivity prediction of DUNE if they are made using two different generators(i.e. GENIE and GiBUU). The dependence of physics results on the selection of the generators will restrict the physics analysis goals.

Figure 4(right panel) shows the mass hierarchy sensitivity study when inverted hierarchy is considered to be true hierarchy. In this analysis, we observe a difference of around 5$\sigma$ for reference beams
and around 6$\sigma$ for optimized beams in the sensitivity results performed by GENIE and GiBUU. This difference is observed in the range $0<\delta_{CP}/\pi<1$ around $\delta_{CP} \sim 0.5/\pi$. In the range $-1<\delta_{CP}/\pi<0$, we observe a variation of around 3$\sigma$ for reference beams and more than 3$\sigma$ for optimized beams at $\delta_{CP} \sim -0.5/\pi$ by GENIE and GiBUU.

\subsection{OCTANT SENSITIVITY}
\begin{figure}
 \centering\includegraphics[scale=.5]{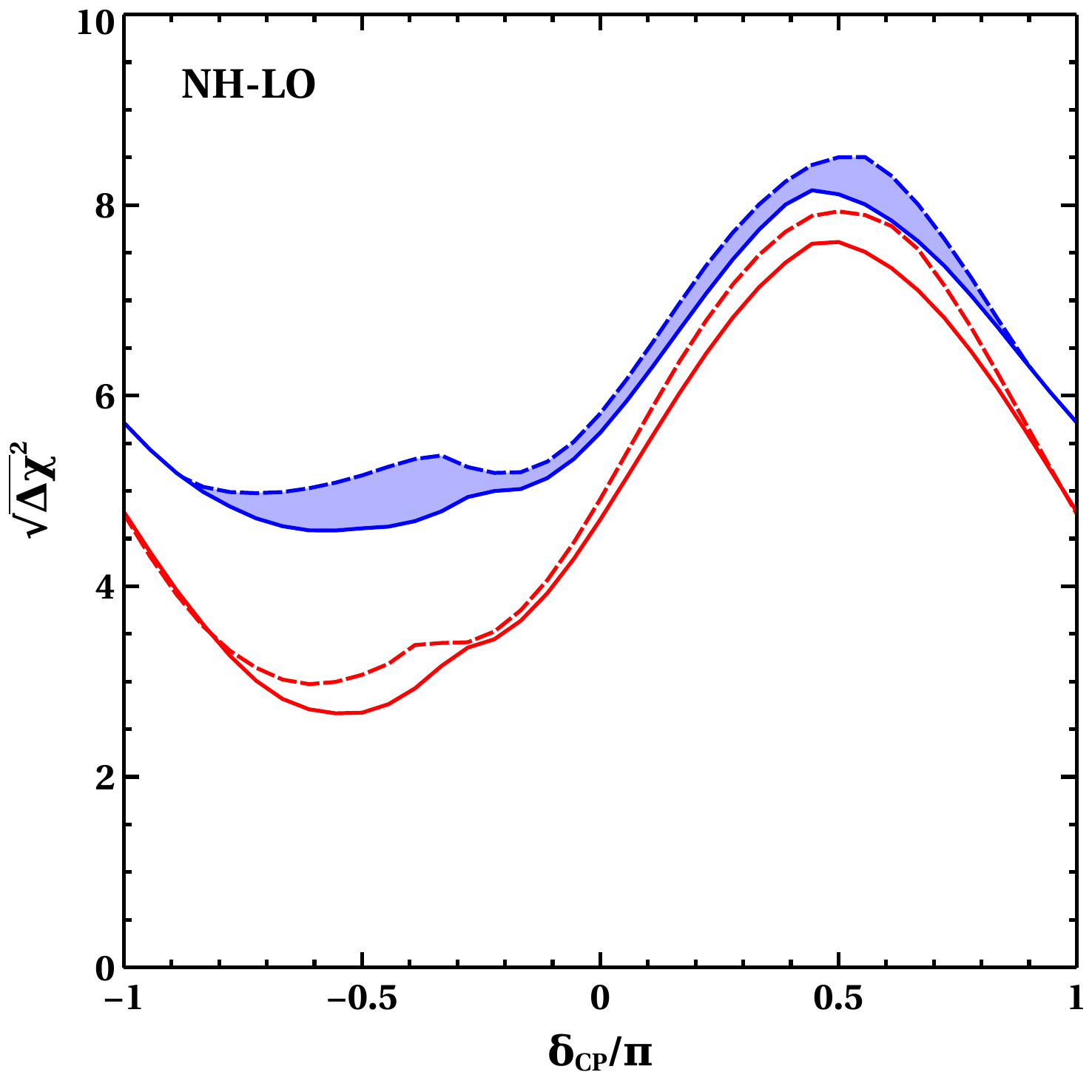}
 \centering\includegraphics[scale=.5]{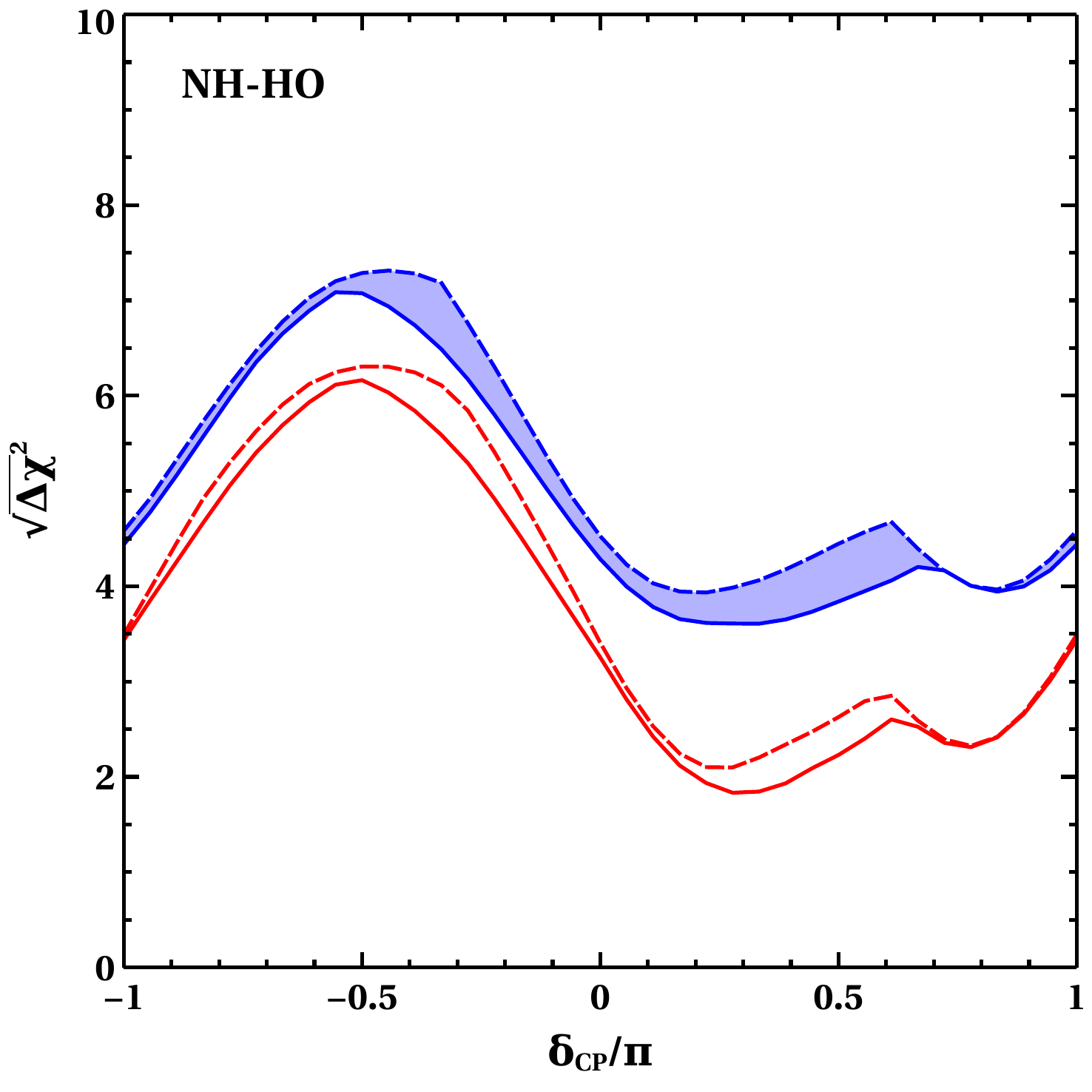}
 \centering\includegraphics[scale=.5]{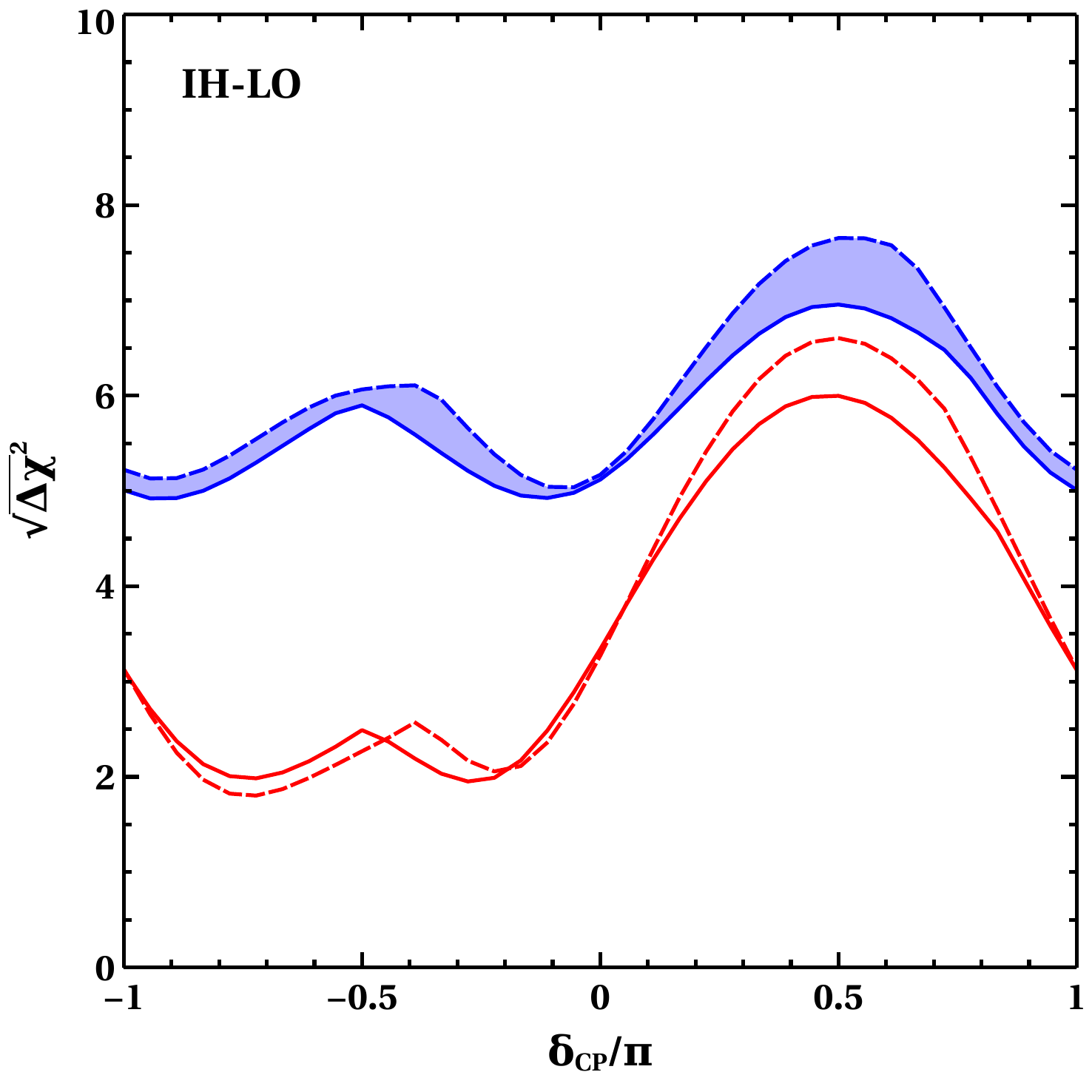}
 \centering\includegraphics[scale=.5]{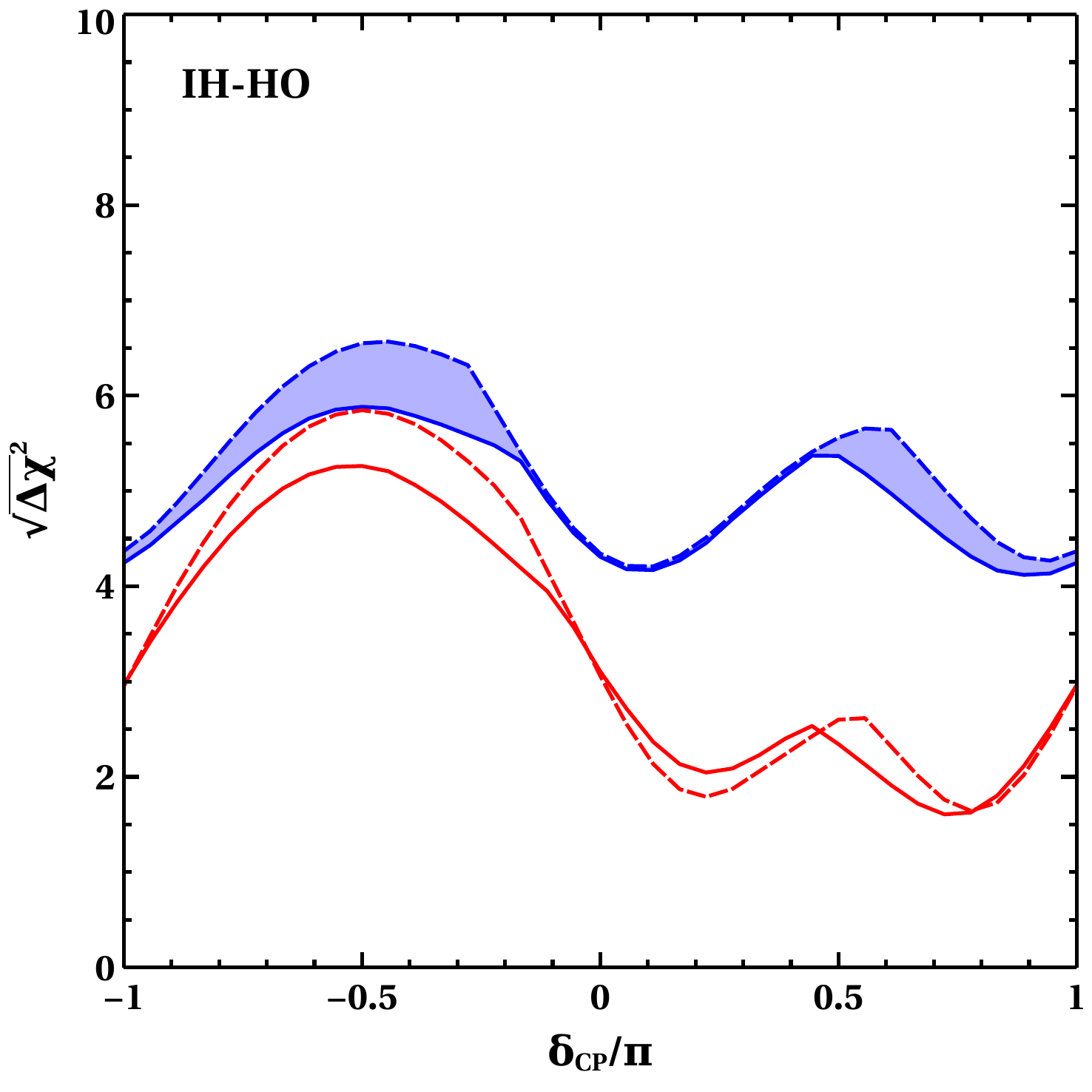}
 \caption{Octant sensitivity measurement as a function of true value of $\delta_{CP}$ for NH-LO(top left), NH-HO(top right), IH-LO(bottom left) and IH-HO(bottom right) by GENIE(blue lines) and GiBUU(red lines)
 Reference and optimized designs are represented by solid and dashed lines respectively.}
\end{figure}
It is not yet confirmed that the atmospheric mixing angle $\theta_{23}$ lies in the lower octant($0<\theta_{23}<\pi/4$)-LO or in the higher octant($\pi/4<\theta_{23}<\pi/2$)-HO with $\pi/4$ as its maximal value.
The prime problem in resolving the octant degeneracy is the appearance of several disconnected regions in the multi-dimensional neutrino oscillation parameter space. 
This makes it difficult to pinpoint the exact or the true solution for a given set of true values, known as parameter degeneracy.
While performing the octant sensitivity calculations in the lower(higher) octant, the test value of $\theta_{23}$ is varied in the lower(higher) octant range. The true value of $\theta_{23}$ in lower octant is 
0.703 and in the higher octant it is 0.867 whereas the range of test values for $\theta_{23}$ in LO is [0.785, 0.961] and for HO is [0.609, 0.785].
For octant sensitivity, the metric $\Delta\chi^{2}$ is defined as-
\begin{equation}
 \Delta\chi^{2}_{octant} = |\chi^{2}_{\theta_{23}^{test}>\pi/4} - \chi^{2}_{\theta_{23}^{true}<\pi/4}|
\end{equation}
We present the results for reference and optimized beam designs for four combinations of true hierarchy and octant configurations viz. NH-LO, NH-HO, IH-LO and IH-HO presented in Figure 5,
top and bottom panels respectively.
For NH-LO case (top left panel of Figure 5), in the negative range of $\delta_{CP}$ values at $\sim-0.5/\pi$ there is a difference of around 2$\sigma$ between sensitivity results by GENIE and GiBUU
and a negligible difference is seen at $\sim0.5/\pi$ in the positive half range of $\delta_{CP}$ values. For NH-HO case(top right panel of Figure 5), the octant sensitivity predictions for reference and optimized
beam designs vary by less than 2$\sigma$ for $\delta_{CP}$ value around 0.5/$\pi$ in the positive range of $\delta_{CP}$ i.e. $0<\delta_{CP}/\pi<1$. A variation of less than 1$\sigma$ is observed in the negative
half range of $\delta_{CP}$ values i.e. $-1<\delta_{CP}/\pi<0$, around $-0.5/\pi$ for both reference and optimized beam designs. For IH-LO case(bottom left panel of Figure 5), there is a substantial 
difference of around 3$\sigma$ between GENIE and GiBUU predictions for reference and optimized beams in the negative half range of $\delta_{CP}$ values, particularly around -0.5/$\pi$ while a difference
of around 1$\sigma$ is observed in the positive half range of $\delta_{CP}$ values between GENIE and GiBUU predictions. For IH-HO case(bottom right panel of Figure 5), 
a difference of less than $\sim3\sigma$ is observed around $0.5/\pi$ and a variation of less than 1$\sigma$ is observed between 
GENIE and GiBUU sensitivity predictions in the negative half range of $\delta_{CP}$ values.

\section{Summary and Conclusion}
Construction of a nuclear model requires the precise combination of information about the energy dependence of all exclusive cross-sections and nuclear effects. 
The simulated results depend on this nuclear model where ignorance of theoretical uncertainties cost inaccuracy. In order to evaluate results from the neutrino
oscillation experiments, we use event generators, which are build upon these nuclear models. They predict the neutrino-nucleon event rates for a particular nuclear 
target along with the topology of final state particles which is critical for oscillation analysis. Due to the use of heavy targets, nuclear effects play an important 
role in the prediction of neutrino oscillation physics. Since nuclear effects are not well understood, thus different generators use different approximations to
accommodate the nuclear effects giving rise to different results.

Our analysis of nuclear effects in the present work reflects the extent of ambiguity in the physics of the selected neutrino event generators. The inbuilt theoretical models in the event generators used in this work i.e. GENIE and GiBUU must be revisited so that the results are independent of the selection of the generators. Since the DUNE flux peaks at 2.5 GeV and at this energy the resonance process is the most dominant process, hence it should be tackled properly. Neutrinos having sufficient energy, strike the nucleon in the nucleus and produce baryonic resonances $N^{*}$, which further quickly decays to a nucleon and a single pion. These are classified as resonant single pion production processes. At lower energies, resonance process is dominated by $\Delta$(1232) production. Contribution from CC $\pi$ production creates complication in neutrino energy determination for disappearance experiments. Thus accurate modeling of nuclear effects in the pion production process is necessary and to mention the largest background contribution to $\nu_{\mu} \rightarrow \nu_{e}$ oscillation channel comes from NC $\pi^{0}$ production.

There is a need to determine cross-section for various nuclei and upgrade the present cross-section data as there is insufficient cross-section data available to understand the existing nuclear effects. The nuclear structure(mass number, atomic number) of every target nucleus is different and poses a non-negligible complication in precise calculation of neutrino-nucleon cross-section. 

The upcoming promising experiment DUNE is designed to answer CP sensitivity, mass hierarchy, octant sensitivity and new physics. To check the potential of DUNE, using
any of the present generators, at the initial step needs a deep understanding and study of the generator itself. Any uncertainty in the generators will be propagated
to the results. Here, DUNE potential is studied using two different generators: GiBUU and GENIE. The analysis performed by different generators for sensitivity studies 
of DUNE is not the same, as can be seen in our analysis(from Figures 3,4 and 5). The variation of few sigmas is observed in the results of two generators.
These uncertainties need to be monitored and addressed while stating the DUNE potential. In our work, only cross-sectional uncertainty is taken into account. 
In a future work, we will consider uncertainties arising from final state interactions while checking the DUNE potential.

\section{Acknowledgment} 
We would like to thank Prof. Raj Gandhi for various suggestions regarding this work. SN and JD would like to thank Prof. U. Mosel for many helpful discussions regarding GiBUU related issues and to 
Newton Nath, Suprabh Prakash, Debajyoti Dutta and Mehedi Masud for their help in GLoBES. This work is supported by the Department of Physics, Lucknow University. Financially it is supported by the Government of India, DST Project no-SR/MF/PS02/2013, Department of Physics, University of Lucknow.

\appendix

Here we present the DUNE sensitivity plots for CP and octant sensitivity in the case of unknown hierarchy. These are represented by figures A1 and B1 respectively.

\section{CP VIOLATION SENSITIVITY}
The CP sensitivity for DUNE is checked for unknown mass hierarchy. The sensitivity plots are illustrated in Figure A1, where left panel is for normal hierarchy and right panel is for inverted hierarchy.

Left panel of Figure A1 shows CP sensitivity when normal hierarchy is considered as true hierarchy. In the positive half range of $\delta_{CP}$ i.e. $0<\delta_{CP}/\pi<1$, at $\delta_{CP} \sim 0.5/\pi$, a variation of more than 1$\sigma$ in the prediction of CP sensitivity by the two generators is observed.

The right panel of Figure A1 shows CP sensitivity when inverted hierarchy is considered as true hierarchy. The CP sensitivity results with the two different generators show a variation of more than 1$\sigma$ 
in the range $0<\delta_{CP}/\pi<1$ around 0.5/$\pi$ for both reference and optimized beam designs. In the negative half range($-1<\delta_{CP}/\pi<0$) of the CP values, the variation between GENIE and
GiBUU predictions for reference and optimized beam designs is seen to be less than 1$\sigma$ for both normal and inverted hierarchy cases.

\begin{figure}
 \centering\includegraphics[scale=.5]{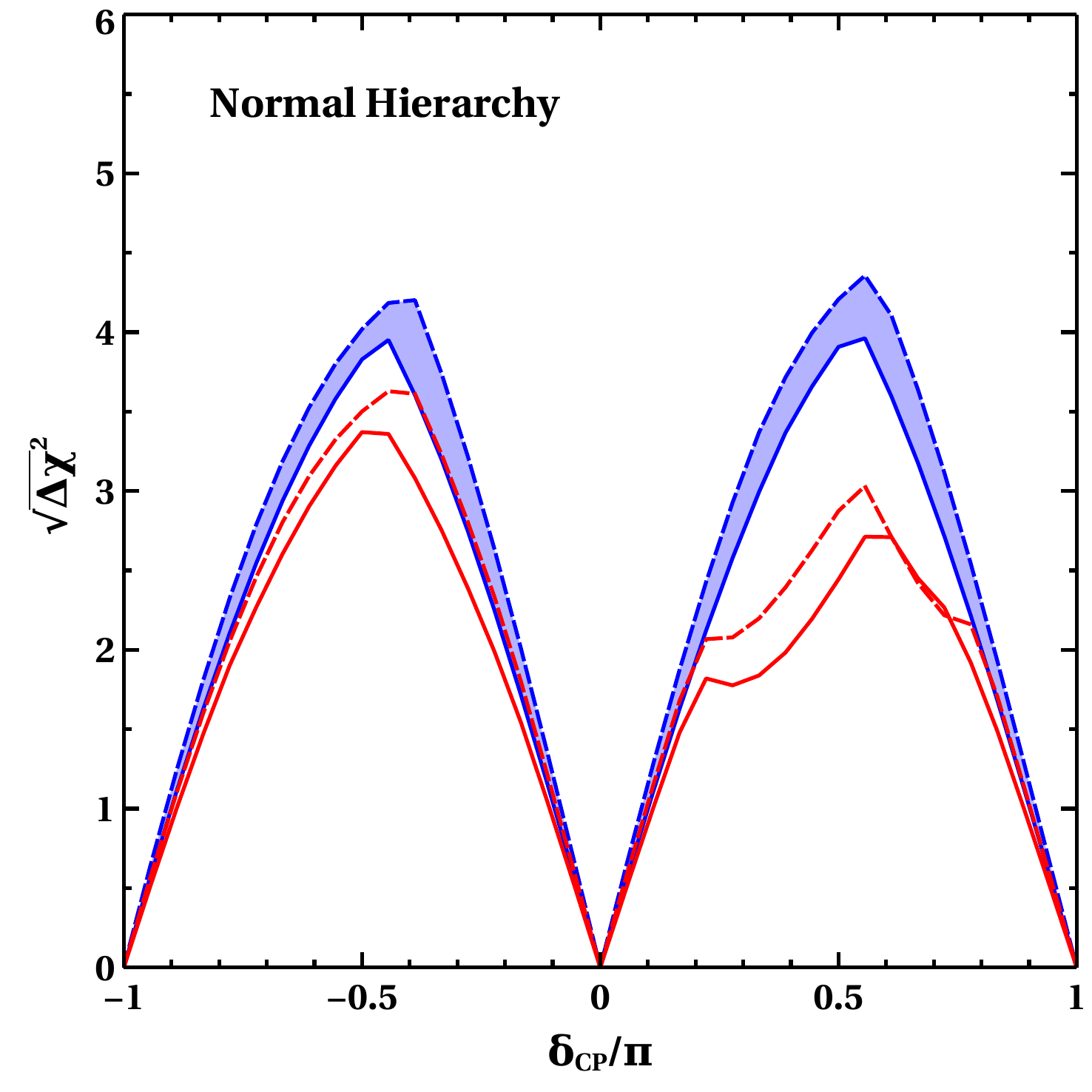}
 \centering\includegraphics[scale=.5]{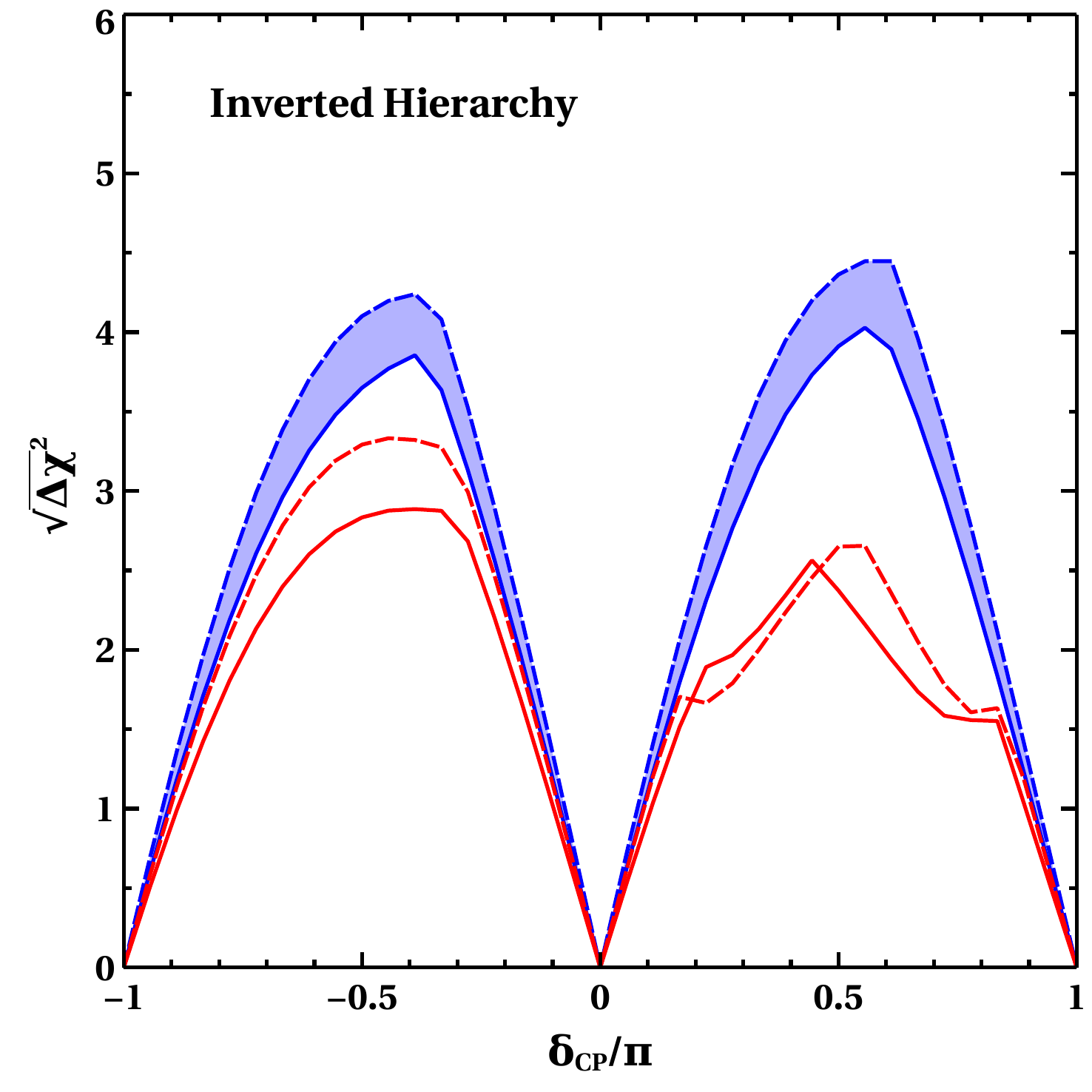}
 \caption{CP sensitivity measurement for unknown hierarchy as a function of true value of $\delta_{CP}$ for NH(left panel) and IH(right panel) by GENIE(blue lines) and GiBUU(red lines). Reference and optimized designs are represented by solid and dashed lines respectively.}
\end{figure}

\section{OCTANT SENSITIVITY}
\begin{figure}
 \centering\includegraphics[scale=.5]{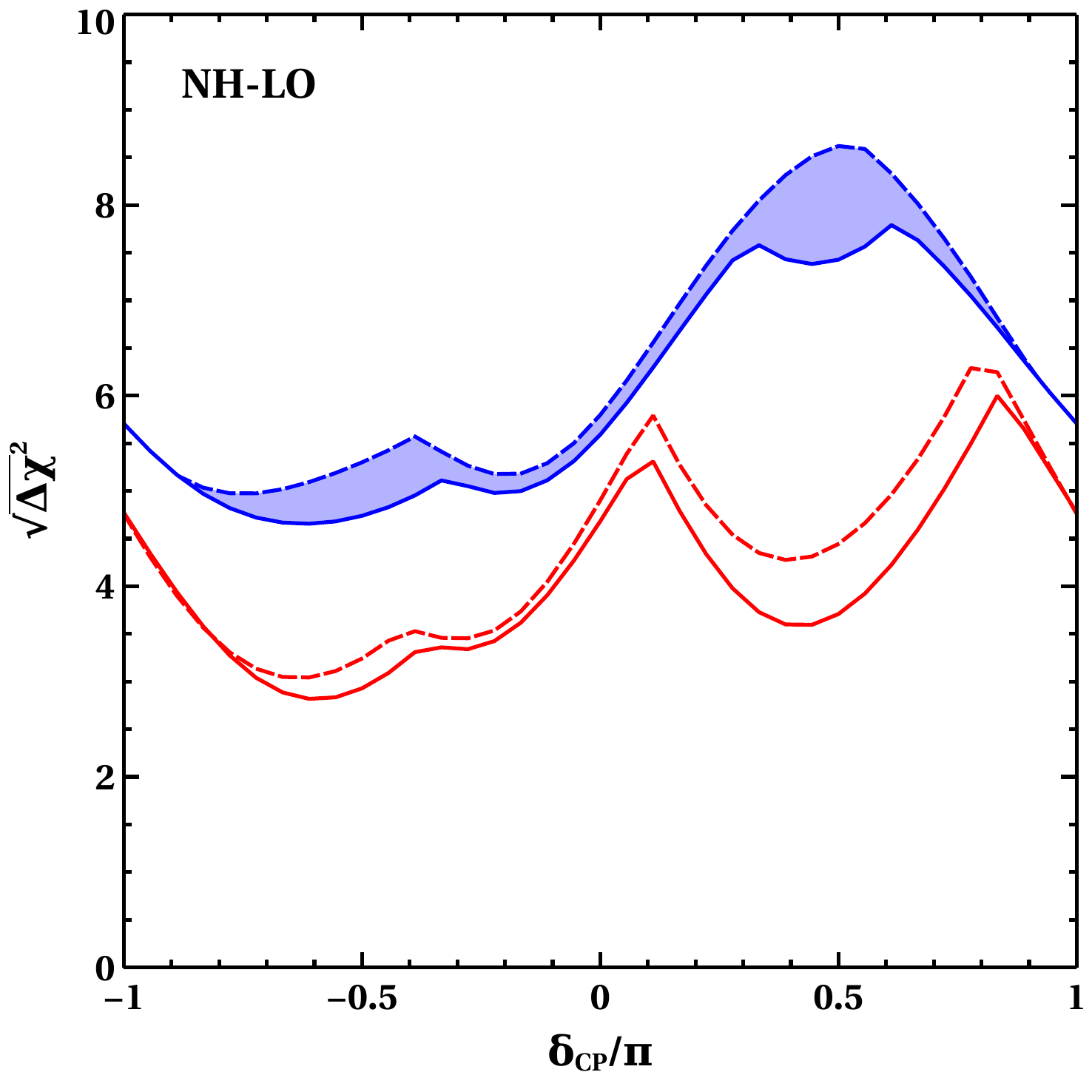}
 \centering\includegraphics[scale=.5]{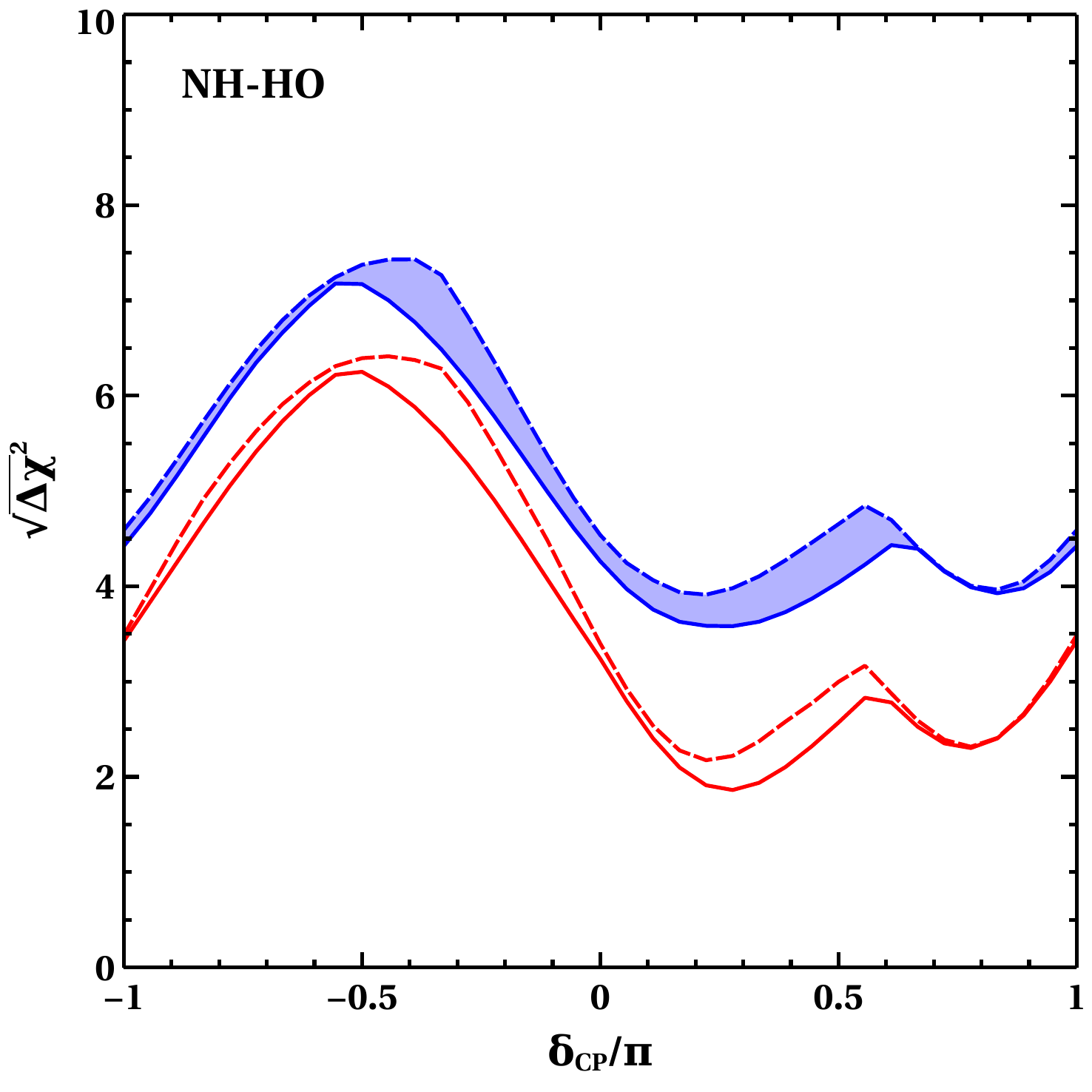}
 \centering\includegraphics[scale=.5]{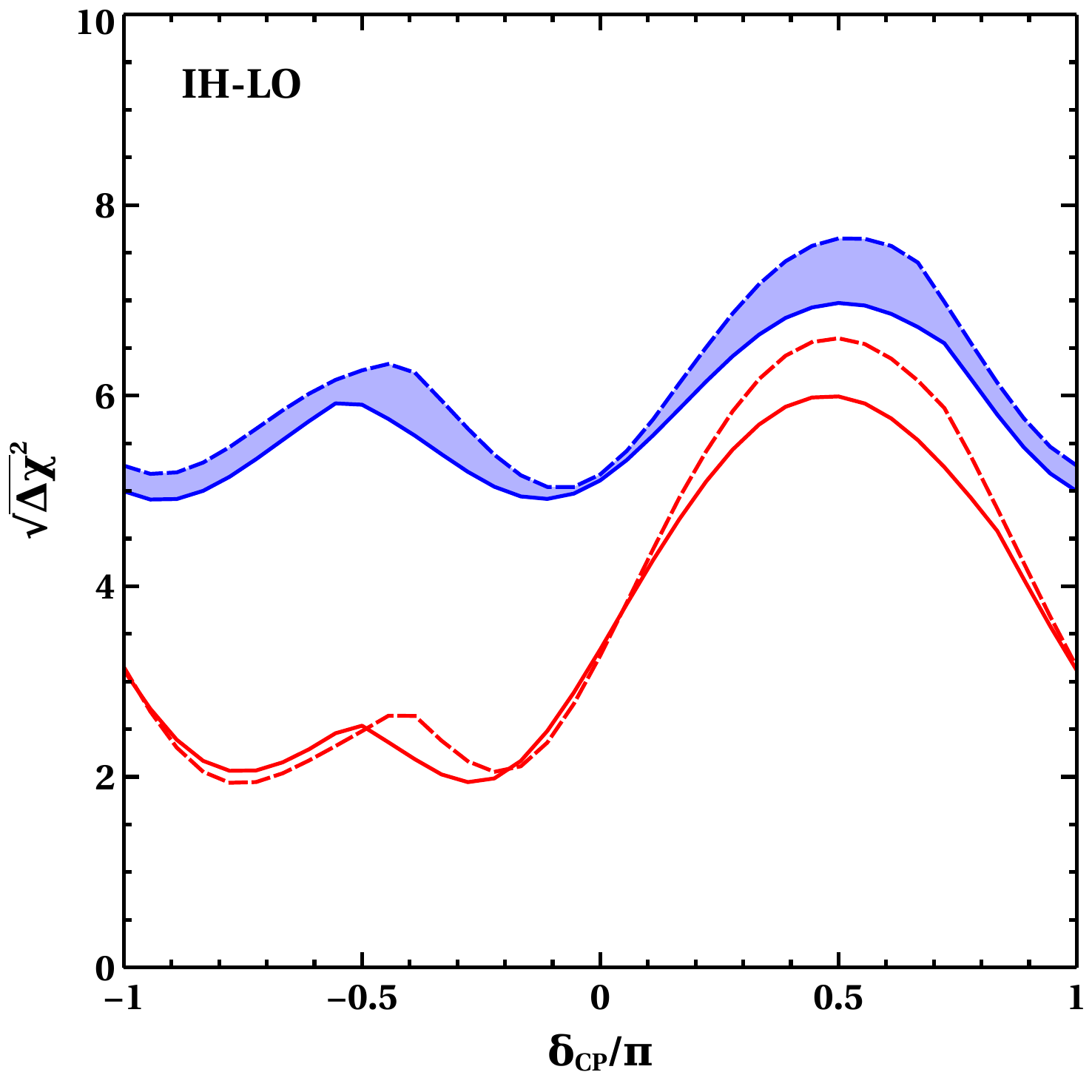}
 \centering\includegraphics[scale=.5]{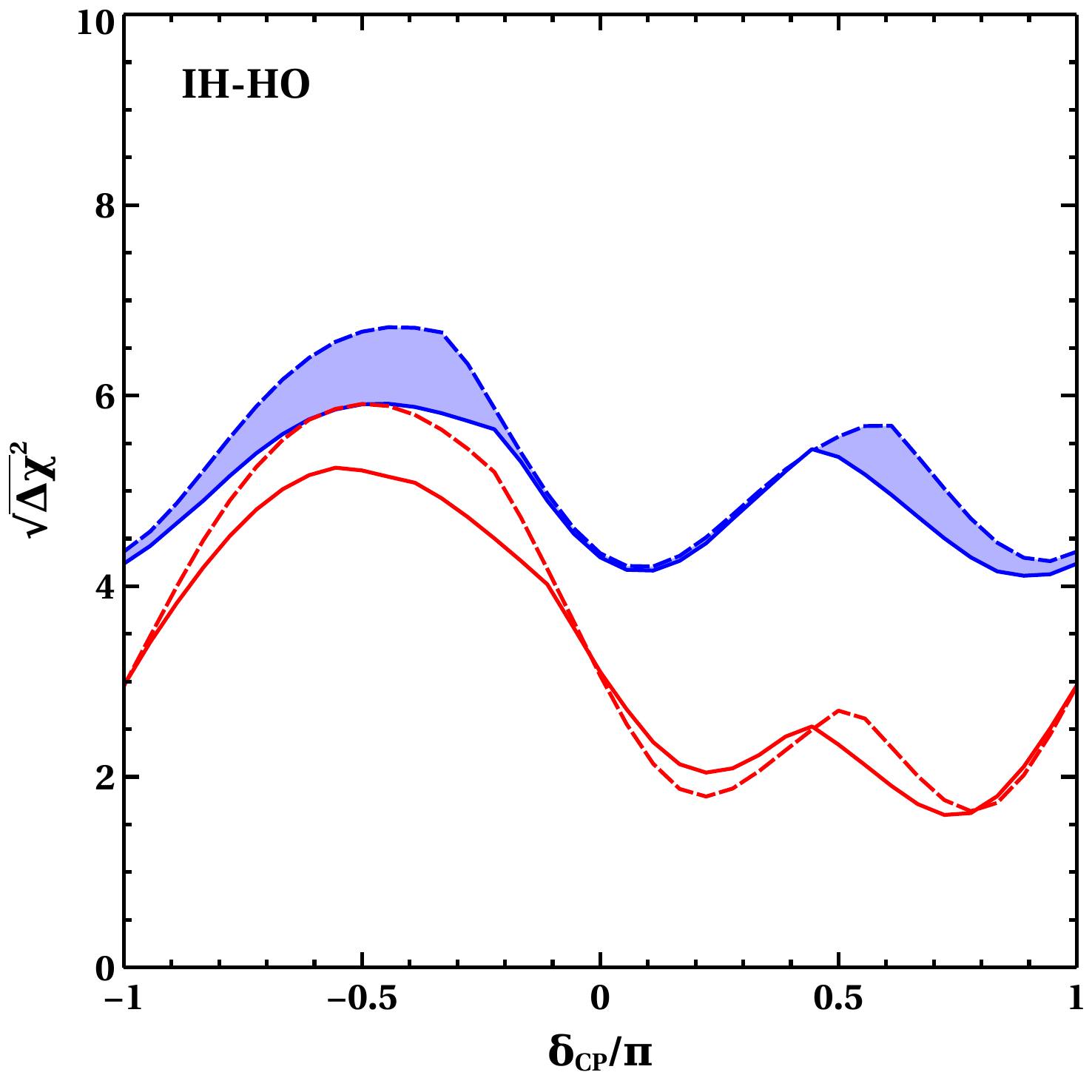}
 \caption{Octant sensitivity measurement for unknown hierarchy as a function of true value of $\delta_{CP}$ for NH-LO(top left), NH-HO(top right), IH-LO(bottom left) and IH-HO(bottom right) by GENIE(blue lines) and GiBUU(red lines) Reference and optimized designs are represented by solid and dashed lines respectively.}
\end{figure}
The octant sensitivity for DUNE is checked for unknown mass hierarchy. The sensitivity plots for reference and optimized beam designs for four combinations of true hierarchy and octant configurations viz. NH-LO, NH-HO, IH-LO and IH-HO are presented in Figure B1, top and bottom panels respectively. 


For NH-LO case (top left panel of Figure B1), in the negative range of $\delta_{CP}$ values at $\sim-0.5/\pi$ there is a difference of around 2$\sigma$ between sensitivity predictions by GENIE and GiBUU
but an appreciable difference of $\sim4\sigma$ is observed at $\sim0.5/\pi$ in the positive half range of $\delta_{CP}$ values. 

For NH-HO case(top right panel of Figure B1), the octant sensitivity predictions for reference and optimized beam designs vary by less than 2$\sigma$ for $\delta_{CP}$ value around 0.5/$\pi$ in the positive range of $\delta_{CP}$ i.e. $0<\delta_{CP}/\pi<1$. A variation of less than 1$\sigma$ is observed in the negative half range of $\delta_{CP}$ values i.e. $-1<\delta_{CP}/\pi<0$, around $-0.5/\pi$ for both reference and optimized beam designs. 

For IH-LO case(bottom left panel of Figure B1), there is a substantial difference of around 3$\sigma$ between GENIE and GiBUU predictions for reference and optimized beams in the negative half range of $\delta_{CP}$ values, particularly around -0.5/$\pi$ while a difference of around 1$\sigma$ is observed in the positive half range of $\delta_{CP}$ values between GENIE and GiBUU predictions. 

For IH-HO case(bottom right panel of Figure B1), a difference of less than $\sim3\sigma$ is observed around $0.5/\pi$ and a variation of less than 1$\sigma$ is observed between GENIE and GiBUU sensitivity predictions in the negative half range of $\delta_{CP}$ values.

\end{document}